\documentclass[preprint,superscriptaddress,onecolumn,showpacs,showkeys]{revtex4}
\usepackage{textcomp}
\usepackage{eurosym}
\usepackage{amsfonts}
\usepackage{array}
\usepackage{amsthm}
\usepackage{bm}
\usepackage{palatino}
\usepackage[colorinlistoftodos]{todonotes}
\usepackage{mathpazo}
\usepackage{supertabular}
\usepackage{subfig}
\usepackage{amssymb}
\usepackage{eurosym}
\usepackage{amsmath}
\usepackage{epsfig}
\usepackage{graphics}
\usepackage{color}
\usepackage{graphicx}
\usepackage[font={footnotesize,it}]{caption}
\usepackage[colorlinks=true,
            linkcolor=red,
            urlcolor=blue,
            citecolor=green]{hyperref}
\usepackage{graphicx,color}

\begin{document}

\title{Analytical Solutions in a Cosmic String Born-Infeld-dilaton Black
Hole Geometry: Quasinormal Modes and Quantization}
\author{\.{I}zzet Sakall\i{}}
\email{izzet.sakalli@emu.edu.tr}
\email{izzet.sakalli@gmail.com}
\affiliation{Physics Department, Faculty of Arts and Sciences, Eastern Mediterranean
University, Famagusta, North Cyprus, via Mersin 10, Turkey}
\author{Kimet Jusufi}
\email{kimet.jusufi@unite.edu.mk}
\affiliation{Physics Department, State University of Tetovo, Ilinden Street nn, 1200,
Tetovo, Macedonia}
\author{Ali \"{O}vg\"{u}n}
\email{ali.ovgun@pucv.cl}
\affiliation{Instituto de F\'{\i}sica, Pontificia Universidad Cat\'olica de Valpara\'{\i}%
so, Casilla 4950, Valpara\'{\i}so, Chile}
\affiliation{Physics Department, Faculty of Arts and Sciences, Eastern Mediterranean
University, Famagusta, North Cyprus, via Mersin 10, Turkey}
\date{\today }

\begin{abstract}
Chargeless massive scalar fields are studied in the spacetime of Born-Infeld
dilaton black holes (BIDBHs). We first separate the massive covariant
Klein-Gordon equation into radial and angular parts and obtain the exact
solution of the radial equation in terms of the confluent Heun functions.
Using the obtained radial solution, we show how one gets the exact
quasinormal modes (QNMs) for some particular cases. We also solve the
Klein-Gordon equation in the spacetime of BIDBH with a cosmic
string in which we point out the effect of the conical deficit on the
BIDBH. The analytical solutions of the radial and angular parts are
obtained in terms of the confluent Heun functions. Finally, we study the
quantization of the BIDBH. While doing this, we also discuss the Hawking
radiation in terms of the effective temperature.
\end{abstract}

\maketitle
\tableofcontents

\setcounter{MaxMatrixCols}{10}


\section{Introduction}

The existence of gravitational waves (GWs), which was confirmed when LIGO
detected GW150914 (stellar-mass binary black holes) in September 2015 (and
announced in February 2016) heralded new era of physics and astronomy \cite%
{TheLIGOScientific:2016wfe}. In this context, QNMs \cite%
{Vishveshwara:1970cc,Vishveshwara:1970zz} of a black hole are related with
the ringdown phase of GWs. Namely, a perturbed black hole settles down
through the characteristic damped oscillations i.e., by the QNMs \cite%
{Chirenti:2017mwe,Blazquez-Salcedo:2016enn,Cardoso:2016rao,Berti:2016lat,Corda:2014ewa}%
. Since the discovery of GWs, studies on the QNMs have gained more attention 
\cite%
{Gonzalez:2017shu,Cruz:2015nza,Gonzalez:2010vv,Lepe:2004kv,Crisostomo:2004hj,Ovgun:2018gwt,Gonzalez:2017zdz,Ovgun:2017dvs,Aros:2002te,Maggiore:2007nq,Wang:2001tk,Konoplya:2003ii,Konoplya:2011qq,Hod:1998vk,Hod:2005dc,Birkandan:2001vr,Birkandan:2017rdp,Fernando:2003ai,Fernando:2003wc,Graca:2016cbd}%
. Moreover, in the gauge/gravity dualities theories, there is a relation
between the QNMs and the poles of a propagator in the dual field theory so
that physicists use it as a tool to work on strong coupled gauge theories
(or holography) \cite%
{ah,Sheykhi:2008rt,Andrade:2017jmt,Andrade:2016rln,Kuang:2016edj,Kovtun:2005ev,Chen:2009hg}%
. 
The frequencies of QNMs are obtained by applying perturbation to the
spacetime of the black hole with appropriate boundary conditions. One of the
most difficulties in studying QNMs is that they admit an eigenvalue problem,
which is not self-adjoint: the system is non-conservative and therefore
energy is lost both at spatial infinity and at the black hole horizon. Thus,
unlike the ordinary normal modes, QNMs are not a basis (see \cite{Berti} and
references therein). Moreover, QNMs can be also used for testing the no-hair
theorem, which states that BHs have only mass, angular momentum and charge
(neutral to quantum effects) and black hole quantization \cite%
{bek,Maggiore:2007nq}. Namely, studying the QNMs can give us new hints about
physics beyond the Einstein's theory of general relativity \cite{gib}.

On the other hand, the complete theory of quantum gravity is still an open
problem. However, black holes can be used as a test arena for the quantum
gravity. From this point of view, Hawking \cite%
{Bekenstein:1974ax,Hawking:1974rv,Hawking,Unruh:1976db,Parikh:1999mf,Akhmedov:2006pg,Kerner:2006vu,Angheben:2005rm,Singleton:2011vh,Kubiznak:2012wp,Kubiznak:2016qmn}
made a pioneering work by showing how the strong gravitational field around
a black hole can affect the production of virtual particles (pairs of
particles and anti-particles: existing all the time in apparently vacuum
according to quantum field theory). According to the theory of Hawking,
virtual particles are likely to be created outside the event horizon of a
black hole. Thus, it is possible that the positive energy (mass) member of
the pair can escape from the black hole - observed as thermal radiation -
while the negative particle (with its negative energy or mass) can fall into
the black hole, which gives rise the black hole to lose its mass. This
phenomenon is known as the Hawking radiation and it was perhaps one of the
first ever examples of the quantum gravity theory. Since then, there is a
continuously growing interest to the thermodynamics of black holes, Hawking
radiation and QNMs \cite%
{Sakalli:2015taa,Sakalli:2014sea,Sakalli:2010yy,Ovgun:2015jna,Sakalli:2016cbo,Sakalli:2017ewb,Jusufi:2015zmv,Jusufi:2015eta,Kuang:2017sqa}%
.

Born and Infeld proposed the non-linear electromagnetic field theory to
solve the problem of divergences in the Maxwell theory \cite{Born:1934ji}.
After the studies on the superstring theory, which admits the D-branes that
are derived from the Born-Infeld action, various researches are done by
using the Einsten-Dilaton-Born-Infeld theory to obtain non-singular charged
black hole solutions. From the high energy levels to the the low energy
limit, the string theory reduces to the Einstein gravity non-minimally
coupled scalar dilaton field \cite{gross,sezgin,dilatonBI,bibh2,bibh3}. The
main motivation for studying the BIDBHs is that they are a promising
candidate for the quantum gravity. In the light of all mentioned above, in
this paper, we shall use the black hole solution in the Born-Infeld-Dilaton
with a Liouville-type potential. This kind of Liouville-type potential
appears when one applies a conformal transformation to the low-energy limit
of the string tree level effective action for the massless boson sector and
in sequel write the action in the Einstein frame \cite%
{sezgin,dilatonBI,bibh2,bibh3}. Furthermore, if the dilaton filed is removed
from the action, the action of Einstein-Born-Infeld with $\Lambda $ is
obtained. Use of the dilaton field causes the changes on the asymptotic
behavior of the spacetime and also leads to curvature singularities at
finite radii. Therefore, such black holes can be used in the studies on
AdS/CFT correspondence, where the holography is occurred near the horizon 
\cite{ah}.

Cosmic strings are one dimensional topological defects extensively studied
in the past \cite%
{Vilenkin:1984ib,Vanchurin:2005pa,Polchinski:2007rg,Blanco-Pillado:2017oxo,Blanco-Pillado:2017rnf,Jusufi:2017uhh}%
. It is widely believed that such topological defects may have formed during
a symmetry breaking process which is usually associated with the phase
transition in the early universe. There are, unfortunately, no experimental
measurements supporting their existence until today. However the possible
detection of these exotic objects in the future will have a great impact in
current understanding on the physics of the early universe. Our main aim in
this paper is to explore an analytical solution of the QNMs and Hawking
radiation in spacetime of BIDBHs \cite{dilatonBI} including the effect of
the cosmic string.

The remainder of this paper is arranged as follows. In Sec. II, we briefly
review the BIDBH spacetime and its thermodynamics. In Sec. III, we study the
Klein-Gordon equation (KGE) of the massive scalar particles propagating in
the geometry of BIDBH. Section IV is devoted to the exact solution of the
radial equation. We also derive the QNMs of BIDBH. In Sec. V, we introduce a
thin cosmic string in the BIDBH spacetime and solve the KGE for a massive
scalar field for that geometry. Then, we obtain the effect of the cosmic
strings on the QNMs of BIDBH spacetime. In Sec. VI, we calculate the
effective temperature and reveal the effects of the cosmic strings on the
quantization of the BIDBH spacetime. In Sec. VII, we summarize our results.

\section{Spacetime of BIDBH}

In this section, we shall review and give the insights of the BIDBH
spacetime \cite{dilatonBI}. The action of the Einstein-dilaton-Born-Infeld
theory, which comprises the dilaton field $\phi $ and the Born-Infeld
parameter $\widehat{\gamma }$ coupled to the Maxwell field is given by 
\begin{equation}
S=\int d^{4}x\sqrt{-g}\Bigl[\mathcal{R}-2(\nabla \phi )^{2}-V(\phi )\ +4%
\widehat{\gamma }e^{-2\tau \phi }\left( 1-\sqrt{1+\frac{F_{\mu \nu }F^{\mu
\nu }}{2\widehat{\gamma }}}\right) \Bigl]  \label{iz1}
\end{equation}%
where $V(\phi )$ denotes potential, $\tau $ is the dilaton coupling
constant, $\mathcal{R}$ stands for the Ricci scalar, and $F_{\mu \nu }$ is
the Maxwell tensor. Dilaton has the following solution \cite{dilatonBI} 
\begin{equation}
\phi (r)=\frac{\tau }{1+\tau ^{2}}\ln (b_{1}r-b_{2})  \label{iz2}
\end{equation}%
where $b_{1}$ and $b_{2}$ are integration constants. Throughout the paper,
we set, without loss of generality, $b_{1}=1$ and $b_{2}=0$. The dilaton
potential can be considered either as $V(\phi )=0$ or as a Liouville-type
potential $V(\phi )=2\Lambda e^{-2\tau \phi }$ [$\Lambda $ is the mass scale
of $V(\phi )$]. For both potentials, the metric function $h(r)$ is found to
be linear in $r$ \cite{bibh2,bibh3}. In other words, $h(r)$ is irrespective
of the dilaton potential and it admits the following solution:

\begin{equation}
h(r)=\frac{r}{L}-b_{0},  \label{iz3}
\end{equation}

where $b_{0}$ is a constant. Length scale $L$ is given by 
\begin{equation}
L^{-1}=2(1-2\rho -\Lambda ),  \label{iz4}
\end{equation}%
where the constant $\rho $ is given by 
\begin{equation}
\rho =-\widehat{\gamma }+\sqrt{\widehat{\gamma }(Q^{2}+\widehat{\gamma })},
\label{iz5}
\end{equation}%
by which $Q$ is the background charge:%
\begin{equation}
Q^{2}=\frac{1+\sqrt{1+16\widehat{\gamma }^{2}}}{8\widehat{\gamma }}.
\label{iz6}
\end{equation}

Finally, as it can be seen from \cite{dilatonBI,bibh2,bibh3}, the
line-element of the BIDBH spacetime takes the following form 
\begin{equation}
ds^{2}=-h(r)dt^{2}+h(r)^{-1}dr^{2}+rd\Omega ^{2}.  \label{iz7}
\end{equation}

Setting $r_{H}=Lb_{0},$ metric function \ref{iz3} recasts in

\begin{equation}
h(r)=\frac{r-r_{H}}{L}.  \label{iz8}
\end{equation}

Thus, one can easily interpret $r_{H}$ as the event horizon of the BIDBH. On
the other hand, due to the non-asymptotically flat character of the metric %
\ref{iz7}, one can employ the Brown-York formalism \cite{byform} to compute
the quasi-local mass, $M_{QL}$ (the interested reader is referred to \cite%
{PRLqlm,fabris} and references therein for $M_{QL}$) of the black hole,
which results in:

\begin{equation}
r_{H}=4LM_{QL}.  \label{iz82}
\end{equation}

\subsection{Thermodynamics of BIDBH}

Considering the following 4-velocity

\begin{equation}
\boldsymbol{u}=u^{t}\partial _{t},  \label{iz9}
\end{equation}

one can easily verify that the normalization condition is satisfied by

\begin{equation}
1=u^{\mu }u_{\mu },  \label{iz10}
\end{equation}

with

\begin{equation}
u^{t}=\frac{1}{\sqrt{g_{tt}}}.  \label{iz11}
\end{equation}

Since the metric components are functions of $r$ and $\theta $, particle
acceleration $a_{p}^{\mu }$ is obtained from \cite{swald}

\begin{equation}
a^{\mu }=-g^{\mu \upsilon }\partial _{\upsilon }\ln u^{t}.  \label{iz12}
\end{equation}

Surface gravity ($\kappa $) is defined by \cite{swald} as follows

\begin{equation}
\kappa =\lim_{r\rightarrow r_{h}}\frac{\sqrt{a^{\mu }a_{\mu }}}{u^{t}},
\label{iz13}
\end{equation}

which yields

\begin{equation}
\kappa =\frac{1}{2}\left. \frac{dh\left( r\right) }{dr}\right\vert
_{r=r_{h}}=\frac{1}{2L}.  \label{iz14}
\end{equation}

Thus, we obtain the Hawking temperature as

\begin{equation}
T_{H}=\frac{\kappa }{2\pi }=\frac{1}{4\pi L}.  \label{iz15}
\end{equation}

It can be deduced from the above result that Hawking radiation of BIDBH\ is
nothing but a isothermal process. Surface area of the BIDBH can be computed
as

\begin{equation}
A_{BH}=\int_{0}^{2\pi }d\varphi \int_{0}^{\pi }\sqrt{-g}\ d\theta =4\pi
r_{h}.  \label{iz16}
\end{equation}%
by which $\sqrt{-g}$ is considered for the hypersurface of BIDBH. Hence, the
entropy of BIDBH becomes \cite{bek,bekent,bekent2} 
\begin{equation}
S_{BH}=\frac{A_{BH}}{4 }=\pi r_{h}.  \label{iz17}
\end{equation}%
The physical quantities given in Eqs. (\ref{iz82}), (\ref{iz16}), and (\ref%
{iz17}) satisfy the first law of thermodynamics: 
\begin{equation}
dM_{QL}=T_{H}dS_{BH}.  \label{iz18}
\end{equation}

\section{QNMs of BIDBH spacetime}

In this section, we shall study the wave equation of the massive scalar
particles propagating in the geometry of BIDBH. To this end, we first
consider the KGE for a massive scalar particle \cite{mkge} 
\begin{equation}
\frac{1}{\sqrt{-g}}\partial _{\alpha }\left( \sqrt{-g}g^{\alpha \nu
}\partial _{\nu }\Psi \right) -m_{b}^{2}\Psi =0,  \label{iz19}
\end{equation}%
where $m_{b}$ is the mass of the boson having scalar field $\Psi $. Owing to
the spherical symmetry and time independence of the spacetime, the scalar
field can be written as 
\begin{equation}
\Psi =\Psi (\boldsymbol{r},t)=P(r)A(\theta )\mbox{e}^{im\varphi }\mbox{e}%
^{-i\omega t}\ ,  \label{iz20}
\end{equation}%
where $\omega $ is the frequency and $m$ denotes the azimuthal quantum
number. Thus, KGE (\ref{iz19}) takes the following form in the BIDBH geometry

\begin{equation*}
-{\frac{{\omega }^{2}r}{h\left( r\right) }}-\frac{{\cot }\left( \theta
\right) }{A\left( \theta \right) }\frac{dA\left( \theta \right) }{d\theta }-%
\frac{1}{A\left( \theta \right) }\frac{d^{2}A\left( \theta \right) }{d{%
\theta }^{2}}+{\frac{{m}^{2}}{\sin ^{2}\left( \theta \right) }}+m_{b}^{2}r-
\end{equation*}

\begin{equation}
{\frac{h\left( r\right) \left[ \frac{dP\left( r\right) }{dr}+r\frac{%
d^{2}P\left( r\right) }{d{r}^{2}}\right] +r\left( {\frac{dh\left( r\right) }{%
dr}}\right) {\frac{dP\left( r\right) }{dr}}}{P\left( r\right) }=0.}
\label{iz21}
\end{equation}

Thus, if one uses an eigenvalue $\lambda ,$ we can separate Eq. (\ref{iz21})
and get an angular equation: 
\begin{equation}
-\frac{1}{A\left( \theta \right) }\frac{d^{2}A\left( \theta \right) }{d{%
\theta }^{2}}-\frac{{\cot }\left( \theta \right) }{A\left( \theta \right) }%
\frac{dA\left( \theta \right) }{d\theta }+{\frac{{m}^{2}}{\sin ^{2}\left(
\theta \right) }+\lambda }=0,  \label{iz22}
\end{equation}%
and a radial equation: 
\begin{equation}
m_{b}^{2}r-{\frac{{\omega }^{2}r}{h\left( r\right) }-\frac{h\left( r\right) %
\left[ \frac{dP\left( r\right) }{dr}+r\frac{d^{2}P\left( r\right) }{d{r}^{2}}%
\right] +r\left( {\frac{dh\left( r\right) }{dr}}\right) {\frac{dP\left(
r\right) }{dr}}}{P\left( r\right) }-\lambda =0.}  \label{iz23}
\end{equation}

As it is well-known, we obtain Legendre polynomials \cite{sf1,sf2} for the
angular equation (\ref{iz22}) with ${\lambda =-l}\left( l+1\right) $ in
which $l$ denotes orbital or azimuthal quantum number.

\subsection{Exact solution of the radial equation and QNMs}

Changing the independent variable from $r$ to $y$ 
\begin{equation}
r=r_{h}\left( 1-y\right) ,  \label{iz24}
\end{equation}%
we transform Eq. (\ref{iz23}) into 
\begin{equation*}
y\left( 1-{y}\right) {\frac{d^{2}P\left( y\right) }{d{y}^{2}}}+\left(
1-2\,y\right) {\frac{dP\left( y\right) }{dy}}+
\end{equation*}

\begin{equation}
\left[ {L}^{2}{\omega }^{2}\left( -1+\frac{1}{y}\right)
)+m_{b}^{2}Lr_{h}\left( 1-y\right) +{l}\left( l+1\right) L\right] P\left(
y\right) =0.  \label{iz25n}
\end{equation}

We also introduce a new function $U(y)$:%
\begin{equation}
P\left( y\right) ={y}^{i\omega \,L}U\left( y\right) ,  \label{iz26}
\end{equation}%
and thus Eq. (\ref{iz25n}) becomes

\begin{equation*}
y\left( 1-{y}\right) {\frac{d^{2}U\left( y\right) }{d{y}^{2}}}+\left[
1-2\,y+2\,i\omega \,L\left( 1-y\right) \right] {\frac{dU\left( y\right) }{dy}%
}+
\end{equation*}

\begin{equation}
\left[ -i\omega \,L+m_{b}^{2}Lr_{h}\left( 1-y\right) +{l}\left( l+1\right) L%
\right] U\left( y\right) =0,  \label{iz27}
\end{equation}

which is the confluent Heun equation \cite{heunc1,heunc2} (and see for
example \cite{heunc3,heunc4,heunc5,heunc6,heunc7,heunc8,heunc9} for its
applications). Its generic form is given by 
\begin{equation}
{\frac{d^{2}U\left( y\right) }{d{y}^{2}}}+\left( \widetilde{a}+\frac{1+%
\widetilde{b}}{y}-\frac{1+\widetilde{c}}{1-y}\right) {\frac{dU\left(
y\right) }{dy}}+\left( \frac{\widetilde{m}}{y}-\frac{\widetilde{n}}{1-y}%
\right) U\left( y\right) =0.  \label{iz28}
\end{equation}

Comparing Eq. (\ref{iz27}) with Eq. (\ref{iz28}), we get

\begin{equation}
\widetilde{a}=\widetilde{c}=0,\ \ \widetilde{b}=2i\omega \,L,  \label{iz29}
\end{equation}

\begin{equation}
\widetilde{m}=\frac{1}{2}(\widetilde{a}-\widetilde{b}-\widetilde{c}+%
\widetilde{a}\widetilde{b}-\widetilde{b}\widetilde{c})-\widetilde{e}=L\left[ 
{l}\left( l+1\right) +m_{b}^{2}r_{h}\right] -i\omega \,L,  \label{iz30}
\end{equation}

\begin{equation}
\widetilde{n}=\frac{1}{2}(\widetilde{a}+\widetilde{b}+\widetilde{c}+%
\widetilde{a}\widetilde{c}+\widetilde{b}\widetilde{c})+\widetilde{d}+%
\widetilde{e}=L\left[ i\omega -{l}\left( l+1\right) \right] .  \label{iz31}
\end{equation}

Thus, we have

\begin{equation}
\widetilde{e}=-L\left[ {l}\left( l+1\right) +m_{b}^{2}r_{h}\right] ,\ \ 
\widetilde{d}=m_{b}^{2}Lr_{h},  \label{iz32}
\end{equation}

The solution of Eq. (\ref{iz28}) is given by \cite{heunc1,heunc2}

\begin{equation}
U\left( y\right) =A_{1}\mbox{HeunC}(\widetilde{a},\widetilde{b},\widetilde{c}%
,\widetilde{d},\widetilde{e};y)+A_{2}y^{-\widetilde{b}}\mbox{HeunC}(%
\widetilde{a},-\widetilde{b},\widetilde{c},\widetilde{d},\widetilde{e};y),
\label{iz33}
\end{equation}

where $A_{1},A_{2}$ are the constants. Thus, the general exact solution of
the radial equation (\ref{iz25n}), in the entire range $-\infty <y\leq 0,$
is given by 
\begin{equation}
P(y)=A_{1}{y}^{i\omega \,L}\mbox{HeunC}(\widetilde{a},\widetilde{b},%
\widetilde{c},\widetilde{d},\widetilde{e};y)+A_{2}{y}^{-i\omega \,L}%
\mbox{HeunC}(\widetilde{a},-\widetilde{b},\widetilde{c},\widetilde{d},%
\widetilde{e};y).  \label{iz34}
\end{equation}

For matching the near horizon and asymptotic regions, we are interested in
the large $r$ behavior ($y\rightarrow -\infty $) of the solution (\ref{iz34}%
). For this purpose, one needs such a connection formula: 
\begin{widetext}
\begin{equation}
\mbox{HeunC}\left( \widetilde{a},\widetilde{b},\widetilde{c},\widetilde{d},%
\widetilde{e};y\right) \rightarrow \text{Gamma function multiplier}\times %
\mbox{HeunC}(\overline{a},\overline{b},\overline{c},\overline{d},\overline{e}%
;y^{-1}),  \label{iz35n}
\end{equation}
\end{widetext}thus the normalization condition \cite{heunc1}: $\mbox{HeunC}(%
\widetilde{a},\widetilde{b},\widetilde{c},\widetilde{d},\widetilde{e}%
;y^{-1}=0)=1$ is going to be satisfied while $y\rightarrow \infty $. The
parameters $\overline{a},\overline{b},\overline{c},\overline{d},\overline{e}$
and the parameters of "Gamma function multiplier" seen in the above equation
have to be related with $\widetilde{a},\widetilde{b},\widetilde{c},%
\widetilde{d},\widetilde{e}$ according to the transformation rules of the
special functions \cite{sf1}. Thus, the asymptotic solution of the radial
equation (\ref{iz34}) would be obtained. In sequel, we should impose the
second boundary condition (pure outgoing QNMs survive at spatial infinity)
and compute the QNMs, analytically. Unfortunately, the absence of Eq. (\ref%
{iz35n}) like transformation in the literature does not allow us to find
analytical forms of the ingoing and outgoing waves at spatial infinity.
Namely, the connection formula (\ref{iz35n})\ of the confluent Heun
functions remained intact \cite{heunc1,heunc2}.

From Eq. (\ref{A12}), the confluent Heun functions can be reduced to the
Gauss hypergeometric functions:

\begin{equation}
\mbox{HeunC}(0,\pm \widetilde{b},0,\widetilde{d},\widetilde{e};y)=\left(
1-y\right) ^{-X_{\pm }}{F\Bigg(X_{\pm },X_{\pm };\,1\pm \widetilde{{b}};\,{%
\frac{y}{y-1}}}\Bigg),  \label{iz36n}
\end{equation}%
if $\widetilde{d}=0\text{ and }1\pm \widetilde{b}\neq 0\text{ for }y\neq 1.$
In Eq. (\ref{iz36n}), ${X_{\pm }}$ are given by

\begin{eqnarray}
X_{\pm } &=&\frac{1\pm \widetilde{b}+\sqrt{\widetilde{{b}}^{2}-4\widetilde{e}%
+1}}{2},  \notag \\
&=&\pm i\omega \,L+\widehat{p},  \label{iz37n}
\end{eqnarray}

in which 
\begin{equation}
\widehat{p}=\frac{1}{2}+i{\omega \Omega },  \label{iz38n}
\end{equation}

where

\begin{equation}
{\Omega =}\sqrt{{L}^{2}-\frac{l\left( l+1\right) L+\frac{1}{4}}{{\omega }^{{2%
}}}}.  \label{iz39n}
\end{equation}

The condition of $\widetilde{d}=0$\ corresponds to the case of massless
bosons ($m_{b}=0$). Therefore, the radial solution (\ref{iz34}) for the
massless scalar fields in the BIDBH\ geometry becomes

\begin{eqnarray}
P(y)&=& A_{1}{y}^{i\omega \,L}\left( 1-y\right) ^{-{X_{+}}}{F\Bigg(%
X_{+},X_{+};1+}\widetilde{{b}}{;\,{\frac{y}{y-1}}}\Bigg)  \notag \\
&+&A_{2}{y}^{-i\omega \,L}\left( 1-y\right) ^{-{X_{-}}}{F\Bigg(X_{-},X_{-};1-%
}\widetilde{{b}}{;\,{\frac{y}{y-1}}}\Bigg),  \label{iz40n}
\end{eqnarray}

Furthermore, if we change the independent variable from $y$ to a new
variable $z$ via the following transformation

\begin{equation}
z={{\frac{y}{y-1}=}}\frac{r-r_{h}}{r},  \label{iz41n}
\end{equation}

the generic \textit{massless} radial solution to Eq. (\ref{iz25n}) reads

\begin{equation}
P(z)=\left( 1-z\right) ^{\widehat{p}}\Big[ A_{1}z^{i\omega \,L}(1-z)^{%
\widehat{p}}F({X_{+},X_{+};1+}\widetilde{{b}};z)+A_{2}z^{-i\omega \,L}F({%
X_{-},X_{-};1-}\widetilde{{b}};z)\Big].  \label{iz42n}
\end{equation}

Tortoise coordinate of the BIDBH is given by

\begin{eqnarray}
r_{\ast } &=&\int \frac{dr}{h\left( r\right) }=L\ln \left( r-r_{h}\right) 
\notag \\
&=&L\int \frac{dy}{y}=L\ln \left( y\right) ,  \label{iz43n}
\end{eqnarray}

It is worth noting that the range $r_{h}<r<\infty $ corresponds to $-\infty <r_{\ast }<\infty $,
since $r_{\ast }\rightarrow -\infty $ as $r\rightarrow r_{h}$. For this
reason, $r_{\ast }$ is known as a tortoise coordinate: as we approach the
horizon, $r$ changes more and more slowly with $r_{\ast }$ since $\frac{dr}{%
dr_{\ast }}\rightarrow 0.$ On the other hand, from Eq. (\ref{iz43n}) we have

\begin{equation}
y=e^{\frac{r_{\ast }}{L}}\text{ \ }\rightarrow \text{ \ }z=\frac{e^{\frac{%
r_{\ast }}{L}}}{e^{\frac{r_{\ast }}{L}}-1}.  \label{iz44n}
\end{equation}

Near the event horizon [$r\rightarrow r_{h}$; $r_{\ast }\rightarrow -\infty $%
; $y\cong z\rightarrow 0$], the hypergeometric functions approximate to one.
Therefore, the radial solution reduces to 
\begin{equation}
P_{NH}(z)\sim A_{1}z^{i\omega \,L}+A_{2}z^{-i\omega \,L}  \label{iz45n}
\end{equation}%
and thus we have the near horizon form of the scalar field $\varphi $ as
follows 
\begin{equation}
\Psi \sim A_{1}e^{-i\omega (t-r_{\ast })}+A_{2}e^{-i\omega (t+r_{\ast })}.
\label{iz46n}
\end{equation}

It is obvious from Eq. (\ref{iz46n}) that first term $\left(
A_{1}e^{-i\omega (t-r_{\ast })}\right) $ represents the outgoing wave, the
second term $\left( A_{2}e^{-i\omega (t+r_{\ast })}\right) $ corresponds to
the ingoing wave. For having the QNMs, we impose one of the boundary
conditions that only ingoing waves survive at the event horizon. To satisfy
this condition, we simply set $A_{1}$ $=0$. Namely, the radial solution
admitting the QNM is given by

\begin{equation}
P(z)=A_{2}z^{-i\omega \,L}\left( 1-z\right) ^{\widehat{p}}F(\overset{%
\widehat{a}}{\overbrace{{X_{-}}}}{,}\overset{\widehat{b}}{\overbrace{{X_{-}}}%
}{;}\overset{\widehat{c}}{\overbrace{{1-}\widetilde{{b}}}};z).  \label{iz47n}
\end{equation}

Performing the connection formula for $z\rightarrow 1-z$ of the
hypergeometric functions \cite{sf1}: 
\begin{widetext}
\begin{align}
P\left( z\right) & =A_{2}z^{-i\omega \,L}(1-z)^{^{\widehat{p}}}\frac{\Gamma (%
\widehat{c})\Gamma (\widehat{c}-\widehat{a}-\widehat{b})}{\Gamma (\widehat{c}%
-\widehat{a})\Gamma (\widehat{c}-\widehat{b})}F(\widehat{a},\widehat{b};1+%
\widehat{a}+\widehat{b}-\widehat{c};1-z)+  \notag \\
& +A_{2}z^{-i\omega \,L}(1-z)^{\widehat{p}-\widehat{a}-\widehat{b}+\widehat{c%
}}\frac{\Gamma (\widehat{c})\Gamma (\widehat{a}+\widehat{b}-\widehat{c})}{%
\Gamma (\widehat{a})\Gamma (\widehat{b})}F(c-\widehat{a},\widehat{c}-%
\widehat{b};1+\widehat{c}-\widehat{a}-\widehat{b};1-z).  \label{iz48n}
\end{align}

in which

\begin{equation}
\widehat{p}-\widehat{a}-\widehat{b}+\widehat{c}=1-\widehat{p}=\frac{1}{2}-i{%
\omega \Omega }.  \label{iz49n}
\end{equation}

Therefore, the asymptotic form ($r\rightarrow \infty $; $r_{\ast
}\rightarrow \infty $; $z\rightarrow 1$) of the radial solution (\ref{iz48n}%
) is given by 
\begin{equation}
P\left( z\right) \sim A_{2}\sqrt{1-z}\left[ (1-z)^{i{\omega \Omega }}\frac{%
\Gamma (\widehat{c})\Gamma (\widehat{c}-\widehat{a}-\widehat{b})}{\Gamma (%
\widehat{c}-\widehat{a})\Gamma (\widehat{c}-\widehat{b})}+(1-z)^{-i{\omega
\Omega }}\frac{\Gamma (\widehat{c})\Gamma (\widehat{a}+\widehat{b}-\widehat{c%
})}{\Gamma (\widehat{a})\Gamma (\widehat{b})}\right] .  \label{iz50n}
\end{equation}

Correspondingly, the near spatial infinity form of the scalar field becomes

\begin{equation}
\Psi \sim A_{2}\sqrt{1-z}\left[ \underset{\text{Ingoing Wave }}{\underbrace{%
e^{-i\omega \left( t-{\Omega }\right) }\frac{\Gamma (\widehat{c})\Gamma (%
\widehat{c}-\widehat{a}-\widehat{b})}{\Gamma (\widehat{c}-\widehat{a})\Gamma
(\widehat{c}-\widehat{b})}}}+\underset{\text{Outgoing Wave}}{e\underbrace{%
^{-i\omega \left( t+{\Omega }\right) }\frac{\Gamma (\widehat{c})\Gamma (%
\widehat{a}+\widehat{b}-\widehat{c})}{\Gamma (\widehat{a})\Gamma (\widehat{b}%
)}}}\right] .  \label{iz51n}
\end{equation}
\end{widetext}
It is worth noting that in order the waves to be ingoing and
outgoing as shown above, we impose a condition, which is named as the 
\textit{wave type identifier condition}by \cite{grg2018,waveid,waveid2}

\begin{equation}
{\Omega =}\sqrt{{L}^{2}-\frac{l\left( l+1\right) L+\frac{1}{4}}{{\omega }^{{2%
}}}}{\in 
\mathbb{R}
>0}.  \label{iz52n}
\end{equation}

Namely, Eq. (\ref{iz52n}) helps us to distinguish the
advanced and retarded times and thus the ingoing and outgoing waves. Spatial infinity boundary condition of QNMs is conditional on the vanishing
ingoing waves at spatial infinity. In other words, only the outgoing waves
are allowed to propagate at the spatial infinity. To this end, one should
terminate the first term (i.e., the ingoing wave)\textbf{.} To this end, we
use the poles of the gamma functions stand in the denominator of the ingoing
wave term [$\Gamma (\widehat{c}-\widehat{a})$ or $\Gamma (\widehat{c}-%
\widehat{b})$] of Eq. (\ref{iz51n}). It is a well-known fact that the gamma
functions $\Gamma (q)$ have poles when $q=-n$ with $n=0,1,2,...$. Thus, QNMs
of the massless scalar waves of the BIDBH are found out as follows: 
\begin{equation}
\widehat{c}-\widehat{a}=\widehat{c}-\widehat{b}=-n,\text{ \ \ \ \ \ \ \ \ (}%
n=0,1,2,...\text{)}  \label{iz53n}
\end{equation}%
From above, we find out the following QNMs: 
\begin{eqnarray}
\omega _{QNM} &=&-{\frac{i\left[ n\left( n+1\right) -Ll\left( l+1\right) %
\right] }{L\left( 1+2\,n\right) },}  \notag \\
&=&-i2\pi T_{H}\frac{\left[ n\left( n+1\right) -Ll\left( l+1\right) \right] 
}{\left( n+\frac{1}{2}\right) }.  \label{iz54n}
\end{eqnarray}

Stable QNMs should have $\mathit{Im}(\omega _{QNM}<0)$. Therefore, one can
immediately observe from Eq. (\ref{iz54n}) that the stable QNMs are
conditional on

\begin{equation}
n\left( n+1\right) \geq Ll\left( l+1\right) .  \label{iz55n}
\end{equation}

One can immediately ask the existence of unstable modes depending on the
values of $n,$ $L,$ and $l$; see Fig. (1). 
\begin{figure}[!htb]
\centering
\includegraphics[scale=0.6]{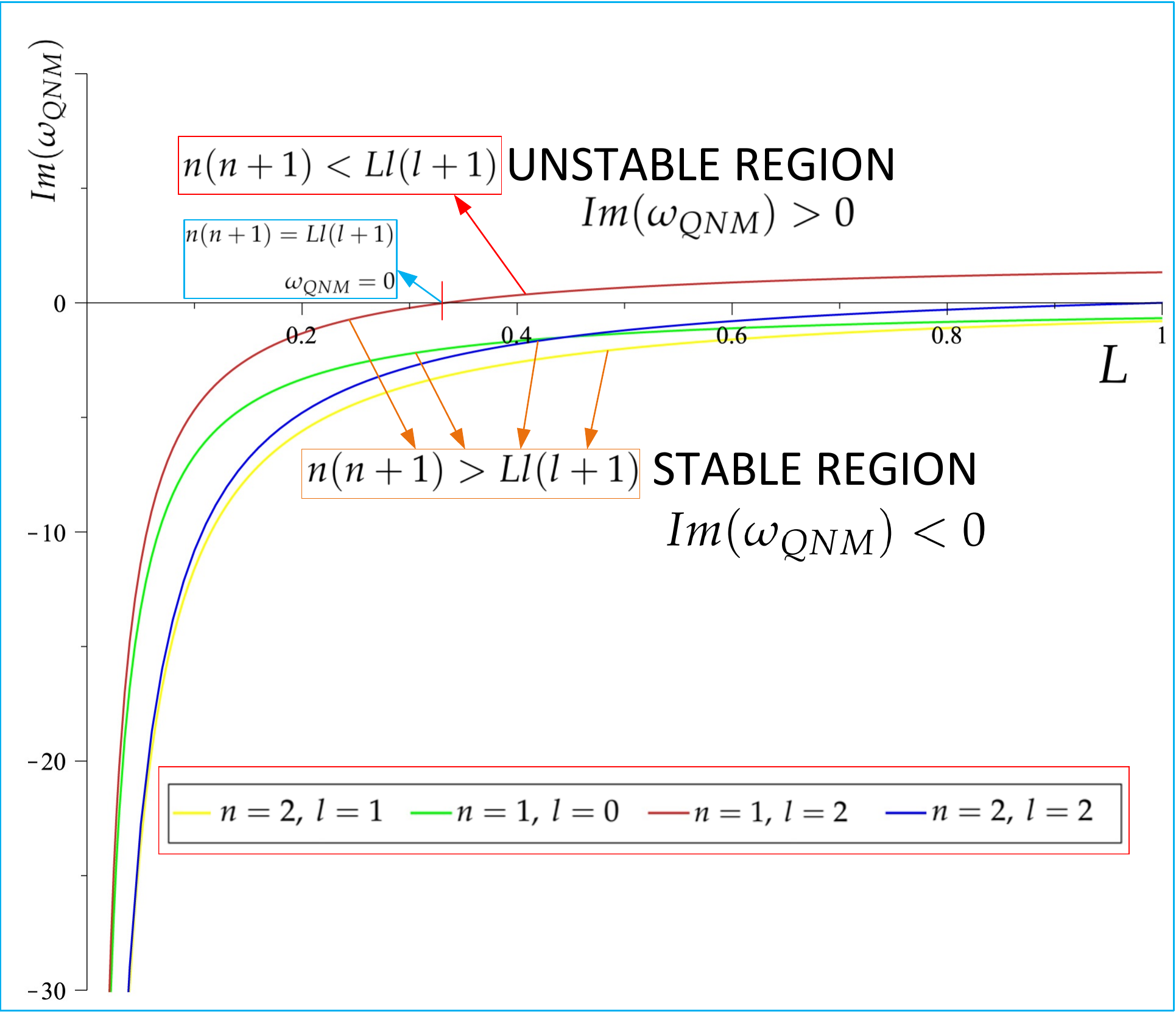}
\caption{Im($\protect\omega _{QNM}$) versus $L$ graph. The plots are
governed by Eq. (\protect\ref{iz54n}). For various $n$ and $l$ values, the
behaviors of $\protect\omega _{QNM}$ are depicted. It is clear that once
condition (\protect\ref{iz55n}) does not hold, the perturbation becomes
unstable \cite{uns1,uns2}) since Im($\protect\omega _{QNM})>0$ shows exponential increase
instead of decay: recall the wave function (\protect\ref{iz20}).}
\label{Figpot}
\end{figure}
However, we want to take the attention of the reader to Eq. (\ref{iz52n}),
which helps us to decide the wave-type: ingoing or outgoing. When $\omega
_{QNM}$ is used in Eq. (\ref{iz52n}), one gets

\begin{equation}
\left. {\Omega }\right\vert _{\omega =\omega _{QNM}}{=}{\frac{\frac{1}{2}%
+n\left( n+1\right) +Ll\left( l+1\right) }{n\left( n+1\right) -Ll\left(
l+1\right) }>0},  \label{iz56n}
\end{equation}%
which is possible only with the condition of Eq. (\ref{iz55n}). It is clear from Eqs. (\ref{iz52n}) and (\ref{iz56n}) that only stable $%
\omega _{QNM}$ allows us to distinguish ingoing and outgoing waves at
spatial infinity. Namely, the analytical method that we applied comprises a
basis for the stable QNMs. 

It is important to remark that while we are about to complete the present
study, we have realized that the problem of QNMs of BIDBH has very recently
studied in \cite{epjc2018}. However, we have differences not only in the
ways to be followed, but in the obtained results as well. We have used the
massive KGE, obtained its radial equation solution in terms of the confluent
Heun functions, and shown how it yields the QNMs of BIDBH in the massless
case. They instead started with massless KGE (relatively simpler scalar wave
equation) and in sequel they directly obtained the hypergeometric function
solution to the radial equation, which made QNMs easily to be found. At the
end, QNM result of \cite{epjc2018} is given by

\begin{eqnarray}
\omega _{n} &=&-i\left( \frac{1}{4L\left( 2n+1\right) }+\frac{l(l+1)}{2n+1}-%
\frac{n+\frac{1}{2}}{2L}\right) ,  \notag \\
&=&{\frac{i\left[ n\left( n+1\right) -Ll\left( l+1\right) \right] }{L\left(
1+2\,n\right) },}  \notag \\
&{=}&{i2\pi T_{H}\frac{\left[ n\left( n+1\right) -Ll\left( l+1\right) \right]
}{\left( n+\frac{1}{2}\right) },}  \label{iz57n}
\end{eqnarray}

which can be easily seen that there is a minus sign difference between Eqs. (%
\ref{iz54n}) and (\ref{iz57n}). In fact, both results are true and they are 
\textit{complementary} to each other. Our result (\ref{iz54n}), which likes
the results of \cite{uns1,our1,our2,Hod:1998vk,our4} admits unstable modes
when $n=0$ but their result (\ref{iz57n}) does not. On the other hand, Eq. (%
\ref{iz57n}) does not admit stability for the $s$-wave ($l=0$) case \cite%
{PRL2004}; QNMs (\ref{iz57n}) lead to exponentially growing modes i.e.,
unstable modes. However, our result (\ref{iz54n}) is stable for the $s$%
-waves, which have strong impact on the testing of the stability since their
motion is perpendicular to the direction of wave propagation.\newline

\section{Spacetime of BIDBH with Cosmic String}

Let us now introduce a thin cosmic string within the BIDBH spacetime. To
this end, letting $\varphi \rightarrow \alpha \varphi $ ($\alpha \in (0,1]$)
for metric (\ref{iz7}), we get 
\begin{equation}
ds^{2}=-h(r)dt^{2}+h(r)^{-1}dr^{2}+rd\theta ^{2}+r\alpha ^{2}\sin ^{2}\theta
d\varphi ^{2},  \label{cs}
\end{equation}%
with $\alpha =1-4\mu $, in which $\mu \geq 0$ is the mass density of the
cosmic string \cite{vieira1}. We may remark here that although the cosmic strings are usually expressed in the cylindrical coordinates for describing the conical topology of the spacetime, for some cases, in particular for the problems in a black hole geometry plus a cosmic string like, it is better to use the spherical coordinates. Namely, the choice of spherical coordinates enables us to make easier computations. After all, no physics changes using different coordinates. In fact, numerous studies in which cosmic strings are expressed in the spherical coordinates are available in the literature (see for instance \cite{csp1,csp2,csp3} and references therein). On the other hand, very recent study of Jose J. Blanco-Pillado 
\cite{Blanco-Pillado:2017rnf} et al shows that the gravitational wave
background to be expected from cosmic strings with the latest pulsar timing
array limits puts an upper bound on the energy scale of the possible cosmic
string network: $\mu<1.5\times10^{-11}$ (in dimensionless units; at $95\%$
confidence level) The surface area and entropy of the BIDBH now modifies to 
\begin{equation}
A_{BH,\alpha }=\int_{0}^{2\pi }d\varphi \int_{0}^{\pi }\sqrt{-g}\ d\theta
=4\pi \alpha r_{h},
\end{equation}

\begin{equation}
S_{BH,\alpha }=\frac{A_{BH}}{4\hslash }=\frac{\pi \alpha r_{h}}{\hslash }.
\end{equation}

Furthermore, since global topology changes, one has to take into account
that the mass (energy) of the system changes this is simply due the volume
change (deficit angle) of the spacetime. To see this, one can calculate the
quasi-local mass \cite{byform} for the spacetime \eqref{cs} as follows 
\begin{equation}
M_{QL,\alpha }=\frac{\alpha }{2}\sqrt{h}\left( \sqrt{h}_{0}-\sqrt{h}\right) ,
\end{equation}%
where 
\begin{equation}
h=\frac{r}{L}-4M_{QL},\,\,\,h_{0}=\frac{r}{L}.
\end{equation}

Using the above equations, in the large radial limit, one can find that 
\begin{equation}
M_{QL,\alpha }=\alpha \,M_{QL}.
\end{equation}

Thus, the first law of thermodynamics is can be written as 
\begin{equation}
dM_{QL,\alpha }=T_{H}\,dS_{BH,\alpha },
\end{equation}%
in which the Hawking temperature Eq. (%
\ref{iz15}) is not affected by the cosmic string.

\section{Exact Solutions and QNMs of BIDBH with Cosmic String}

One can show that, with ansatz Eq. (\ref{iz20}), the radial equation of the
massive KGE on the geometry of Eq. (\ref{cs}) remains intact, however the
angular equation takes the following form 
\begin{equation}
-\frac{1}{A\left( \theta \right) }\frac{d^{2}A\left( \theta \right) }{d{%
\theta }^{2}}-\frac{{\cot }\left( \theta \right) }{A\left( \theta \right) }%
\frac{dA\left( \theta \right) }{d\theta }+{\frac{{m}^{2}}{\alpha ^{2}\,\sin
^{2}\left( \theta \right) }+\lambda }=0,  \label{k70}
\end{equation}%
From Eq. \eqref{k70}, one can naturally set 
\begin{equation}
m_{(\alpha )}=\frac{m}{\alpha },
\end{equation}%
In the case of $\alpha \neq 1$, the quantum number $m_{(\alpha )}$ is no
longer integer. Thus, Eq. \eqref{k70} can be recast in the following form 
\begin{equation}
\frac{1}{\sin \theta }\frac{d}{d\theta }\left( \sin \theta \frac{dA}{d\theta 
}\right) +\left( \lambda -\frac{m_{(\alpha )}^{2}}{\sin ^{2}\theta }\right)
A=0.
\end{equation}

This type of differential equation was exactly solved by \cite%
{vieira1,vieira2} in the following way. First, one needs to introduce a new
coordinate $w=\cos ^{2}\theta $, which transforms Eq. \eqref{k70} to 
\begin{equation*}
\frac{d^{2}A}{dw^{2}}+\left( \frac{1/2}{w}+\frac{1}{w-1}\right) \frac{dA}{dw}%
+
\end{equation*}%
\begin{equation}
\left[ \frac{\lambda -m_{(\alpha )}^{2}}{4}\frac{1}{w}+\frac{m_{(\alpha
)}^{2}-\lambda }{4}\frac{1}{w-1}-\left( \frac{m_{(\alpha )}^{2}}{2}\right)
^{2}\frac{1}{(w-1)^{2}}\right] A=0.  \label{iz58}
\end{equation}

Then, if we let $A(w)=(w-1)^{m_{(\alpha )}/2}V(w)$, the above equation
reduces to the confluent Heun equation: 
\begin{equation}
\frac{d^{2}V}{dw^{2}}+\left( \underline{\tilde{a}}+\frac{\underline{\tilde{b}%
}+1}{w}+\frac{\tilde{c}+1}{w-1}\right) \frac{dV}{dw}+\left( \frac{\tilde{m}}{%
w}+\frac{\tilde{n}}{w-1}\right) V=0,
\end{equation}%
where the parameters of HeunC$(\tilde{a},\tilde{b},\tilde{c},\tilde{d},%
\tilde{e};w)$ now are given by 
\begin{equation}
\tilde{a}=0,\,\tilde{b}=-\frac{1}{2},\,\tilde{c}=m_{(\alpha )},\,\tilde{d}%
=0,\,\tilde{e}=\frac{1+m_{(\alpha )}^{2}-\lambda }{4}.
\end{equation}

The general solution to Eq. \eqref{iz58} thus becomes 
\begin{equation}
A(w)=(w-1)^{m_{(\alpha )}/2}\Big(C_{1}\text{HeunC}(\tilde{a},\tilde{b},%
\tilde{c},\tilde{d},\tilde{e};w)+C_{2}w^{-\tilde{b}}\text{HeunC}(\tilde{a},-%
\tilde{b},\tilde{c},\tilde{d},\tilde{e};w)\Big),
\end{equation}%
in which $C_{1}$ and $C_{2}$ are two integration constants. Hence, we remark
that the general behavior of the scalar wave function (\ref{iz20}) of the
BIDBH is influenced by the presence of the cosmic string $\alpha $.

It is interesting to note that the radial solution is quite similar to the
case covered in the previous section, thus we are going to skip the solution
procedure. However, there is a crucial point that must be considered. In the
existence of the cosmic string, the energy of the particles measured at
spatial infinity changes since the volume of the spacetime has a deficit
angle. That is to say, when a wave packed (particle) is emitted from the
black hole, its energy is shifted by a factor of $\alpha $. For this reason,
we have 
\begin{equation}
\omega \rightarrow \omega _{\alpha }=\omega \alpha .  \label{iz59}
\end{equation}

So, the general wave solution of the BIDBH with cosmic string is given by
[recall Eq. (\ref{iz51n})] 
\begin{widetext}
\begin{equation}
\Psi \sim A_{2}\sqrt{1-z}\left[ \underset{\text{Ingoing Wave }}{\underbrace{%
e^{-i\omega_\alpha \left( t-{\Omega_\alpha }\right) }\frac{\Gamma (\widehat{c})\Gamma (%
\widehat{c}-\widehat{a}-\widehat{b})}{\Gamma (\widehat{c}-\widehat{a})\Gamma
(\widehat{c}-\widehat{b})}}}+\underset{\text{Outgoing Wave}}{e\underbrace{%
^{-i\omega_\alpha \left( t+{\Omega_\alpha }\right) }\frac{\Gamma (\widehat{c})\Gamma (%
\widehat{a}+\widehat{b}-\widehat{c})}{\Gamma (\widehat{a})\Gamma (\widehat{b}%
)}}}\right],  
\end{equation}
\end{widetext}in which

\begin{equation}
{\Omega _{\alpha }=}\sqrt{{L}^{2}-\frac{l\left( l+1\right) L+\frac{1}{4}}{{%
\omega _{\alpha }}^{{2}}}}{\in 
\mathbb{R}
>0}.
\end{equation}%
In the presence of the cosmic string, Eq. (\ref{iz59}) thus modifies the
QNMs (\ref{iz54n}) of the BIDBH as 
\begin{equation}
\omega _{QNM,\alpha }=\omega _{QNM}\,\alpha =-{\frac{i\alpha \,\left[
n\left( n+1\right) -Ll\left( l+1\right) \right] }{L\left( 1+2\,n\right) }}.
\label{plot}
\end{equation}

We remark that the general scalar wave solution is modified when the cosmic
string is introduced and thus the QNMs are shifted by a parameter $\alpha $.
In this way, QNMs might provide a new method to detect topological defects
like cosmic strings by the virtue of GWs.

\section{Effective Temperature and Quantization of BIDBH}

Hawking radiation \cite{swald} is a quantum process associated with the
quantum fields near the event horizon. During the Hawking radiation one
expects not only photons, but also gravitons to be emitted from the black
hole, which, in a classical level, can be viewed as the GW \cite{dejan}.
Since a black hole is an open system (it continuously interacts with the
quantum fields), this implies a loss in the energy via Hawking radiation.
Thus, it is natural to expect a possible link between the QNMs and black
hole thermodynamics, but there is no satisfactory answer yet. In fact, Hod 
\cite{HOD} first argued that the damping rate of the fundamental ($n=0$)
QNMs of any black hole is constrained by its Hawking temperature via 
\begin{equation}
|\text{Im}\omega |\leq \pi \,T_{H}.
\end{equation}

One can check the validity of Hod's conjecture by using the fundamental
modes of the $s$-waves. In this case, Eq. \eqref{iz54n} admits the following
imaginary part 
\begin{equation}
|\text{Im}\omega |=4\pi \,T_{H}>\pi \,T_{H}.  \label{ix1}
\end{equation}%
The result obtained in Eq. (\ref{ix1}) is against to the Hod's conjecture.
However, considering the fact that metric (\ref{cs}) is non-asymptotically
flat, such a result should not be seen as very surprising. Besides, the
violation of Hod's conjecture has been recently reported in the context of
Gauss-Bonnet de Sitter black hole \cite{zhidenko}.

We now consider the asymptotic highly damped ($n\rightarrow \infty $) QNMs %
\eqref{iz54n}, which has the imaginary part related to the Hawking
temperature in a very simple way: 
\begin{equation}
\omega _{n}\simeq -2\,\pi \,i\,T_{H}\,\left( n+\frac{1}{2}\right) =-i\kappa
_{H}\left( n+\frac{1}{2}\right) .  \label{k119}
\end{equation}

We recall that the Parikh-Wilczek's quantum tunneling method \cite%
{Parikh:1999mf}, which includes the back reaction effects in the spacetime
i.e., a change of the black hole horizon when a Hawking quanta is emitted,
this implies a deviation from the thermal nature of the Hawking spectrum.
Corda \cite{corda1,corda2,corda3}, inspiring from the Bohr correspondence
principle and the quantum tunneling method, argued that QNMs can be
re-interpreted by introducing the effective temperature $T_{H}\rightarrow
T_{E}$, which incorporates the back reaction effects in the black hole
horizon. Along this line of thinking, Eq. \eqref{k119} should take the
following form 
\begin{equation}
\omega _{n}\simeq -2\,\pi \,i\,T_{E}\,\left( n+\frac{1}{2}\right) .
\label{xi2}
\end{equation}

As a result of the quantum tunneling near the event horizon, we can
interpret the imaginary part, seen in the above equation, as a loss of the
black hole energy via QNMs. In other words, as it was suggested by Maggiore 
\cite{Maggiore:2007nq}, a black hole behaves pretty much like a quantum
harmonic oscillator. Namely, it constantly interacts with the quantum fields
and hence never stops to oscillate. This argument is also supported by a
recent paper \cite{alberto}, which argues that highly damped modes always
exist and are related to the presence of the horizon.

The probability of emission is given by \cite{Parikh:1999mf} 
\begin{equation}
\Gamma _{BH}=\exp \left[ -\frac{\omega _{QL}}{T_{H}}\left( 1-\omega
_{QL}\right) \right] .
\end{equation}

For this reason, if one wants to take into account the dynamical geometry of
the BH during the emission of the particle, the effective temperature \cite%
{corda1,corda2,corda3} should be considered: 
\begin{equation}
T_{E}\equiv \frac{T_{H}}{1-\omega _{QL}}=\frac{1}{4\pi L\left( 1-\omega
_{QL}\right) }.
\end{equation}

Thus, Eq. (\ref{xi2}) becomes 
\begin{equation}
\omega _{n}\simeq -\frac{\,i}{2L(1-|\omega _{n}|)}\,\left( n+\frac{1}{2}%
\right) ,  \label{k125}
\end{equation}%
where $|\omega _{n}|=E_{n}=\omega _{QL,n}$ is the total energy emitted at
level $n$. We impose a condition that a black hole cannot emit more energy
than their total mass, i.e., $|\omega _{n}|\leq M$. From Eq. \eqref{k125} it
follows that 
\begin{equation}
|\omega _{n}|=\frac{1}{2}+\frac{\sqrt{L-1-2n}}{2\sqrt{L}},
\end{equation}%
which yields the following condition 
\begin{equation}
n\leq n_{max}=\frac{L-1}{2}.
\end{equation}%
Now, if we consider an emission from the ground state (i.e., a BH which is
not excited) to a state with $n=n_{1}$, we can write down 
\begin{equation}
M_{n_{1}}=M-|E_{n_{1}}|=M-\frac{1}{2}-\frac{\sqrt{L-1-2n_{1}}}{2\sqrt{L}},
\end{equation}%
and similarly to a state with $n=n_{2}$, 
\begin{equation}
M_{n_{2}}=M-|E_{n_{2}}|=M-\frac{1}{2}-\frac{\sqrt{L-1-2n_{2}}}{2\sqrt{L}}.
\end{equation}

Taking the difference of these two energy states, one gets 
\begin{eqnarray}
\Delta E_{n_{1}\rightarrow n_{2}} &=&M_{n_{2}}-M_{n_{1}},  \notag \\
&=&-\frac{\sqrt{L-1-2n_{2}}}{2\sqrt{L}}+\frac{\sqrt{L-1-2n_{1}}}{2\sqrt{L}}.
\end{eqnarray}

Now setting $n_{1}=n-1$ and $n_{2}=n$, we find out 
\begin{equation*}
\Delta E_{n_{1}\rightarrow n_{2}}=\frac{\sqrt{L-2n+1}}{2\sqrt{L}}-\frac{%
\sqrt{L-2n-1}}{2\sqrt{L}}.
\end{equation*}

It can be assumed that the length scale is the same for emission and
absorption. Let us define the effective length scale $L_{E(n,n-1)}$ as \cite%
{corda3} 
\begin{equation}
L_{E(n,n-1)}=\frac{L_{n-1}+L_{n}}{2}=\frac{\sqrt{L-2n-1}}{2\sqrt{L}}+\frac{%
\sqrt{L-2n+1}}{2\sqrt{L}}.
\end{equation}%
Since $A_{BH}=16\pi LM$, the change in area can be defined as 
\begin{equation}
\Delta A_{BH}=16\pi L_{E(n,n-1)}\Delta E_{n_{1}\rightarrow n_{2}},
\end{equation}

in which

\begin{equation*}
L_{E(n,n-1)}\Delta E_{n_{1}\rightarrow n_{2}}=\frac{1}{2L},
\end{equation*}

and thus we have 
\begin{equation}
\Delta A_{BH}=8\pi .  \label{Fin}
\end{equation}%
Equation \eqref{Fin} is nothing but the original result of Bekenstein \cite%
{bek}. Note that this relation seems to be universal and not affected when
the Hawking radiation spectrum is non-thermal \cite{corda1,corda2}. We also
remark that the Hod's conjecture is not only violated, but also affected by
the conical topology, when we replace $T_{H}\rightarrow T_{E}$; because the
imaginary part of the fundamental modes reads 
\begin{equation}
|\text{Im}\omega _{\alpha }|=4\,\pi \,T_{E}(1-4\mu )>\pi T_{E}.
\end{equation}

On the other hand, one can also show the validity of the Bekenstein's result 
\cite{bek} by employing the method of Maggiore \cite{Maggiore:2007nq}. For
the highly damping modes ($n\rightarrow \infty $), the transition frequency
can be obtained from Eq. \eqref{k119} as follows 
\begin{equation}
\Delta \omega \approx \text{Im}\left( \omega _{n-1}-\omega _{n}\right)
=\kappa _{H}=\frac{1}{2L}.
\end{equation}

With the help of the adiabatic invariance formula: 
\begin{equation}
I_{adb}=\int \frac{T_{H}\Delta S_{BH}}{\Delta \omega },
\end{equation}%
which basically suggests the Bohr-Sommerfeld quantization condition: $%
I_{adb,n\rightarrow \infty }=n\hbar $, we find out that entropy is quantized
as follows 
\begin{equation}
S_{BH,n}=2\pi n.
\end{equation}

Thus, the area spectrum can be easily found as 
\begin{equation}
A_{BH,n}=4\hbar S_{BH,n}=8\pi \hbar n.
\end{equation}

One can observe that this reveals the Bekenstein's conjecture $%
[A_{n}=\epsilon n\hbar ]$, with the minimum change to the area of the
horizon i.e., $\Delta A_{min}=A_{n}-A_{n-1}=8\pi \hbar $, or $\Delta
A_{min}=8\pi $ within the geometric unit sytem: $\hbar =1$ \cite{bek}.

When considering the cosmic string, the adiabatic invariance formula would
be modified to 
\begin{equation}
I_{adb}=\int \frac{T_{H}\Delta S_{BH,\alpha }}{\Delta \omega _{\alpha }},
\end{equation}

in which

\begin{equation}
\Delta \omega _{\alpha }=\alpha \Delta \omega =\alpha \kappa _{H},
\end{equation}

we get the entropy change as $S_{BH,n\alpha }=2\pi \alpha n$. Hence, $%
A_{BH,n\alpha }=4S_{BH,n\alpha }=8\pi \alpha n$, and the change in area
results in $\Delta A_{min,\alpha }=8\pi \alpha $. In other words, the
minimal change in entropy/area spectrum is affected by the cosmic string
parameter.

\section{Conclusions}

We have studied the analytical solution to the KGE for a massive scalar
field in the BIDBH spacetime. Radial exact solution is given in terms of the
confluent Heun functions \cite{heunc1,heunc2}, and it covers the whole range
of the observable space $0\leq z<1$.

The obtained radial solution could not be extended to the asymptotic region
because of the lack of the inverse connection formula (\ref{iz35n}) which
would help us to get the exact asymptotic form of the radial solution. This
gap has enforced us to consider the massless scalar fields. Using the
particular transformation (\ref{A12}), we have managed to express the
confluent Heun functions in terms of the hypergeometric functions.\ By this
was, we have transferred the radial solution from the near horizon region to
asymptotic region \cite{sf1,sf2}. Afterward, we have computed QNMs of the
massless scalar waves propagating in the BIDBH. In particular, Eqs. (\ref%
{iz55n}) and (\ref{iz56n}) have shown us that our analytical QNM results are
possible with stable waves. We have also compared Eq. (\ref{iz54n}) with the
QNM results of \cite{epjc2018}. We have remarked that both results are
complementary with each other. It is important to note that our solution (%
\ref{iz54n}) violates the Hod's conjecture \cite{Hod:1998vk} such as being
discussed in \cite{zhidenko}.

In the presence of a cosmic string, QNMs are found to be shifted by a
parameter $\alpha $ when a cosmic string is introduced:

\begin{equation}
\omega _{QNM,\alpha }=-{\frac{i\,\left[ n\left( n+1\right) -Ll\left(
l+1\right) \right] }{L\left( 1+2\,n\right) }}\alpha ,
\end{equation}

which obeys the same condition of Eq. (\ref{iz55n}) to possess the stable
QNMs. In short, our QNM analysis might provide an information to identify
the topological defects (i.e., cosmic strings) in the fingerprints of the
GWs. We have also shown that the minimum change in the area of the black
hole is shifted by the cosmic string parameter $\alpha $, which means a
deviation from the Bekenstein's result \cite{bek}: $\Delta A_{min}=8\pi
\alpha $.

The results of the present study motivate us for doing further works in this
direction. In the near future, we plan to extend our analytical analysis to
the other fields (Dirac fields, vector fields, gravitons etc.) and explore
the effects of spin and cosmic string on the QNMs and quantization of the black hole.

\acknowledgments

We wish to thank the Editor and anonymous Referee for their valuable comments
and suggestions. I. S. is grateful to Dr. Edgardo Cheb-Terrab
(Waterloo-Canada) for his valuable comments on the confluent Heun functions.
A. \"{O}.~acknowledges financial support provided under the Chilean FONDECYT
Grant No. 3170035. A. \"{O}. is grateful to Prof. Robert Mann for hosting
him as a research visitor at Waterloo University.

\section{Appendix}

The confluent Heun equations is obtained from the general Heun equation \cite%
{heunc1,app1,app2} through a confluence process, that is, a process where
two singularities coalesce, performed by redefining parameters and taking
limits, resulting in a single (typically irregular) singularity. The
confluent Heun equation is given by

\begin{equation}
{\frac{d^{2}U\left( y\right) }{d{y}^{2}}}+\left( \widetilde{a}+\frac{1+%
\widetilde{b}}{y}-\frac{1+\widetilde{c}}{1-y}\right) {\frac{dU\left(
y\right) }{dy}}+\left( \frac{\widetilde{m}}{y}-\frac{\widetilde{n}}{1-y}%
\right) U\left( y\right) =0.  \tag{A1}  \label{A1}
\end{equation}

and thus \eqref{A1} has three singular points: two regular ones: $y=0$ and $%
y=1$, and one irregular one: $y=\infty $. Solution of \eqref{A1} is called
the confluent Heun's function : $U\left( y\right) =\mbox{HeunC}(\widetilde{a}%
,\widetilde{b},\widetilde{c},\widetilde{d},\widetilde{e};y)$, which is
regular around the regular singular point $y=0$. It is defined as 
\begin{equation}
\mbox{HeunC}(\widetilde{a},\widetilde{b},\widetilde{c},\widetilde{d},%
\widetilde{e};y)=\sum_{n=0}^{\infty }u_{n}(\widetilde{a},\widetilde{b},%
\widetilde{c},\widetilde{d},\widetilde{e})y^{n},  \tag{A2}  \label{A2}
\end{equation}%
which is the convergent in the disk $|y|<1$ and satisfies the normalization $%
\mbox{HeunC}(\widetilde{a},\widetilde{b},\widetilde{c},\widetilde{d},%
\widetilde{e};0)=1$. The parameters $\widetilde{a},\widetilde{b},\widetilde{c%
},\widetilde{d},\widetilde{e}$ are related with $\widetilde{m}$ and $%
\widetilde{n}$ as follows

\begin{equation}
\widetilde{m}=\frac{1}{2}(\widetilde{a}-\widetilde{b}-\widetilde{c}+%
\widetilde{a}\widetilde{b}-\widetilde{b}\widetilde{c})-\widetilde{e}, 
\tag{A3}  \label{A3}
\end{equation}

\begin{equation}
\widetilde{n}=\frac{1}{2}(\widetilde{a}+\widetilde{b}+\widetilde{c}+%
\widetilde{a}\widetilde{c}+\widetilde{b}\widetilde{c})+\widetilde{d}+%
\widetilde{e}.  \tag{A4}  \label{A4}
\end{equation}

The coefficients $u_{n}(\widetilde{a},\widetilde{b},\widetilde{c},\widetilde{%
d},\widetilde{e})$ are determined by three-term recurrence relation: 
\begin{equation}
A_{n}u_{n}=B_{n}u_{n-1}+C_{n}u_{n-2},  \tag{A5}  \label{A5}
\end{equation}%
with initial conditions \{$u_{-1}=0\,$,$\,u_{0}=1$\} and we have 
\begin{equation}
\hskip-.truecmA_{n} =1+{\frac{\widetilde{b}}{n}}\,\rightarrow 1,\,\,\,\text{%
as}\,\,\,n\rightarrow \infty ,  \notag
\end{equation}
\begin{equation}
\hskip-.truecmB_{n} =1+{\frac{-\widetilde{a}+\widetilde{b}+\widetilde{c}-1}{n%
}}+{\frac{\widetilde{e}+(\widetilde{a}-\widetilde{b}-\widetilde{c})/2+%
\widetilde{b}/2\left( \widetilde{c}-\widetilde{a}\right) }{n^{2}}}%
\,\rightarrow 1,\,\,\,\text{as}\,\,\,n\rightarrow \infty ,  \tag{A6}
\label{A6}
\end{equation}
\begin{equation}
\hskip-.truecmC_{n} ={\frac{\widetilde{a}}{n^{2}}}\left( {\frac{\widetilde{d}%
}{\widetilde{a}}}+{\frac{\widetilde{b}+\widetilde{c}}{2}}+n-1\right)
\,\rightarrow 0,\,\,\,\text{as}\,\,\,n\rightarrow \infty .  \notag
\end{equation}

$\mbox{HeunC}(\widetilde{a},\widetilde{b},\widetilde{c},\widetilde{d},%
\widetilde{e};y)$ reduces to a polynomial of degree $N\left(
=0,1,2,....\right) $ with respect to the variable $y$ if and only if the
following two conditions are satisfied \cite{heunc3}:

\begin{equation}
{\frac{\widetilde{d}}{\widetilde{a}}}+{\frac{\widetilde{b}+\widetilde{c}}{2}}%
+N+1 =0,  \tag{A7}  \label{A7}
\end{equation}%
\begin{equation}
\Delta _{N+1}(\widetilde{m})=0.  \tag{A8}  \label{A8}
\end{equation}

\eqref{A7} is known as "$\delta _{N}$-condition" and \eqref{A8} is called "$%
\Delta _{N+1}$-condition". In fact, the $\delta _{N}$-condition is nothing
but $C_{N+2}=0$ and the $\Delta _{N+1}$-condition corresponds to $u_{N+1}(%
\widetilde{a},\widetilde{b},\widetilde{c},\widetilde{d},\widetilde{e})=0$.

Since the confluent Heun equation thus has two regular singularities and one
irregular singularity, it includes the $_{2}F_{1}$ hypergeometric equation 
\cite{sf1}:

\begin{equation}
\left( -z+{z}^{2}\right) {\frac{d^{2}}{d{z}^{2}}}Y\left( z\right) +\left[
\left( 1+\widehat{a}+\widehat{b}\right) z-\widehat{c}\right] {\frac{d}{dz}}%
Y\left( z\right) +\widehat{a}\widehat{b}Y\left( z\right) =0  \tag{A9}
\label{A9}
\end{equation}%
which can be expressed in terms of $\mbox{HeunC}$ functions as follows \cite%
{app2} 
\begin{equation}
Y\left( z\right) =\mathit{C}_{\mathit{1}}\,\left( 1-z\right) ^{-\widehat{a}}%
\mbox{HeunC}\left( 0,\widehat{a}-\widehat{b},\widehat{c}-1,0,\frac{1}{2}\,%
\left[ \left( \widehat{c}-2\,\widehat{a}\right) \widehat{b}-\,\widehat{c}%
\left( 1-\widehat{a}\right) +1\right] ,\left( 1-z\right) ^{-1}\right) + 
\notag
\end{equation}%
\begin{equation}
\mathit{C}_{\mathit{2}}\,\left( 1-z\right) ^{-\widehat{b}}\mbox{HeunC}\left(
0,\widehat{b}-\widehat{a},\widehat{c}-1,0,\frac{1}{2}\,\left[ \left( 
\widehat{c}-2\,\widehat{a}\right) \widehat{b}-\,\widehat{c}\left( 1-\widehat{%
a}\right) +1\right] ,\left( 1-z\right) ^{-1}\right)  \tag{A10}  \label{A10}
\end{equation}%
In fact, the $_{2}F_{1}$ hypergeometric function is related to $\mbox{HeunC}$
function by \cite{app2} 
\begin{equation}
\begin{split}
_{2}F_{1}(\widehat{a},\widehat{b};\,\widehat{c};\,z)& =\left( 1-z\right) ^{-%
\widehat{b}}\mbox{HeunC}\left( 0,\widehat{c}-1,\widehat{b}-\widehat{a},0,%
\frac{1}{2}\,\left( -1+\widehat{a}+\widehat{b}\right) \widehat{c}-\widehat{a}%
\widehat{b}+\frac{1}{2},{\frac{-z}{1-z}}\right) , \\
\:\:\:\text{where}\:\:\:z& \neq 1.
\end{split}
\tag{A11}
\end{equation}

Inversely, $\mbox{HeunC}$ function can be rewritten in terms of the $%
_{2}F_{1}$ hypergeometric function as follows \cite{app2,app3,app4}

\begin{equation}
\begin{split}
\mbox{HeunC}\left( 0,\widehat{b},\widehat{c},0,\widehat{e},z\right) &
=\left( 1-z\right) ^{-\frac{1}{2}\left( 1+\widehat{b}+\widehat{c}+\sqrt{1+%
\widehat{{b}}^{2}+\widehat{{c}}^{2}-4\widehat{e}}\right) }\times \\
& {_{2}F_{1}}\left[ \frac{1}{2}\left( 1+\widehat{b}+\widehat{c}+\sqrt{1+%
\widehat{b}^{2}+\widehat{c}^{2}-4\widehat{e}}\right) \right. ,\, \\
& \left. \frac{1}{2}\left( 1+\widehat{b}-\widehat{c}+\sqrt{1+\widehat{b}^{2}+%
\widehat{c}^{2}-4\widehat{e}}\right) ;\,1+\widehat{b};\,{\frac{z}{-1+z}}%
\right] , \\
\:\:\:\text{where}\:\:\:& 1+\widehat{b}\neq 0\:\:\:\text{and}\:\:\:z\neq 1.
\end{split}
\tag{A12}  \label{A12}
\end{equation}


\begin{thebibliography}{999}
\bibitem{TheLIGOScientific:2016wfe} B.~P.~Abbott \textit{et al.} [LIGO
Scientific and Virgo Collaborations], 
Phys.\ Rev.\ Lett.\ \textbf{116}, no. 24, 241102 (2016).

\bibitem{Vishveshwara:1970cc} C.~V.~Vishveshwara, 
Phys.\ Rev.\ D \textbf{1}, 2870 (1970).

\bibitem{Vishveshwara:1970zz} C.~V.~Vishveshwara, 
Nature \textbf{227}, 936 (1970).

\bibitem{Chirenti:2017mwe} C.~Chirenti, 
Braz.\ J.\ Phys.\ \textbf{48}, no. 1, 102 (2018).

\bibitem{Blazquez-Salcedo:2016enn} J.~L.~Blazquez-Salcedo, C.~F.~B.~Macedo,
V.~Cardoso, V.~Ferrari, L.~Gualtieri, F.~S.~Khoo, J.~Kunz and P.~Pani, 
Phys.\ Rev.\ D \textbf{94}, no. 10, 104024 (2016).

\bibitem{Cardoso:2016rao} V.~Cardoso, E.~Franzin and P.~Pani, 
Phys.\ Rev.\ Lett.\ \textbf{116}, no. 17, 171101 (2016).

\bibitem{Berti:2016lat} E.~Berti, A.~Sesana, E.~Barausse, V.~Cardoso and
K.~Belczynski, 
Phys.\ Rev.\ Lett.\ \textbf{117}, no. 10, 101102 (2016).

\bibitem{Corda:2014ewa} C.~Corda, 
Class.\ Quant.\ Grav.\ \textbf{32}, no. 19, 195007 (2015).

\bibitem{Gonzalez:2017shu} P.~A.~Gonzalez, E.~Papantonopoulos, J.~Saavedra
and Y.~Vasquez, 
Phys.\ Rev.\ D \textbf{95}, no. 6, 064046 (2017).

\bibitem{Cruz:2015nza} M.~Cruz, M.~Gonzalez-Espinoza, J.~Saavedra and
D.~Vargas-Arancibia, 
Eur.\ Phys.\ J.\ C \textbf{76}, no. 2, 75 (2016).

\bibitem{Gonzalez:2010vv} P.~Gonzalez, E.~Papantonopoulos and J.~Saavedra, 
JHEP \textbf{1008}, 050 (2010).

\bibitem{Lepe:2004kv} S.~Lepe and J.~Saavedra, 
Phys.\ Lett.\ B \textbf{617}, 174 (2005).

\bibitem{Crisostomo:2004hj} J.~Crisostomo, S.~Lepe and J.~Saavedra, 
Class.\ Quant.\ Grav.\ \textbf{21}, 2801 (2004).

\bibitem{Ovgun:2018gwt} A.~\"{O}vg\"{u}n and K.~Jusufi, 
arXiv:1801.02555 [gr-qc].

\bibitem{Gonzalez:2017zdz} P.~A.~Gonzalez, A.~\"{O}vg\"{u}n, J.~Saavedra and
Y.~Vasquez, 
arXiv:1711.01865 [gr-qc].

\bibitem{Ovgun:2017dvs} A.~\"{O}vg\"{u}n, I.~Sakalli and J.~Saavedra, 
Chin. Phys. C \textbf{42}, 105102 (2018).

\bibitem{Aros:2002te} R.~Aros, C.~Martinez, R.~Troncoso and J.~Zanelli, 
Phys.\ Rev.\ D \textbf{67}, 044014 (2003).

\bibitem{Maggiore:2007nq} M.~Maggiore, 
Phys.\ Rev.\ Lett.\ \textbf{100}, 141301 (2008).

\bibitem{Wang:2001tk} B.~Wang, E.~Abdalla and R.~B.~Mann, 
Phys.\ Rev.\ D \textbf{65}, 084006 (2002).

\bibitem{Konoplya:2003ii} R.~A.~Konoplya, 
Phys.\ Rev.\ D \textbf{68}, 024018 (2003).

\bibitem{Konoplya:2011qq} R.~A.~Konoplya and A.~Zhidenko, 
Rev.\ Mod.\ Phys.\ \textbf{83}, 793 (2011).

\bibitem{Hod:1998vk} S.~Hod, 
Phys.\ Rev.\ Lett.\ \textbf{81}, 4293 (1998).

\bibitem{Hod:2005dc} S.~Hod, 
Class.\ Quant.\ Grav.\ \textbf{23}, L23 (2006).

\bibitem{Birkandan:2001vr} T.~Birkandan and M.~Hortacsu, 
Gen.\ Rel.\ Grav.\ \textbf{35}, 457 (2003).

\bibitem{Birkandan:2017rdp} T.~Birkandan and M.~Hortacsu, 
EPL \textbf{119}, no. 2, 20002 (2017).

\bibitem{Fernando:2003ai} S.~Fernando, 
Gen.\ Rel.\ Grav.\ \textbf{36}, 71 (2004).

\bibitem{Fernando:2003wc} S.~Fernando and K.~Arnold, 
Gen.\ Rel.\ Grav.\ \textbf{36}, 1805 (2004).

\bibitem{Graca:2016cbd} J.~P.~Morais Graca, G.~I.~Salako and V.~B.~Bezerra, 
Int.\ J.\ Mod.\ Phys.\ D \textbf{26}, no. 10, 1750113 (2017)

\bibitem{ah} O.~Aharony, M.~Berkooz, D.~Kutasov and N.~Seiberg, 
JHEP \textbf{9810}, 004 (1998).

\bibitem{Sheykhi:2008rt} A.~Sheykhi, 
Int.\ J.\ Mod.\ Phys.\ D \textbf{18}, 25 (2009).

\bibitem{Andrade:2017jmt} T.~Andrade, 
arXiv:1712.00548 [hep-th].

\bibitem{Andrade:2016rln} T.~Andrade, J.~Casalderrey-Solana and A.~Ficnar, 
JHEP \textbf{1702}, 016 (2017).

\bibitem{Kuang:2016edj} X.~M.~Kuang and E.~Papantonopoulos, 
JHEP \textbf{1608}, 161 (2016).

\bibitem{Kovtun:2005ev} P.~K.~Kovtun and A.~O.~Starinets, 
Phys.\ Rev.\ D \textbf{72}, 086009 (2005).

\bibitem{Chen:2009hg} B.~Chen and Z.~b.~Xu, 
JHEP \textbf{0911}, 091 (2009).

\bibitem{Berti} E.~Berti, K.~Yagi, H.~Yang and N.~Yunes, 
arXiv:1801.03587 [gr-qc]; "to be appeared in the journal of General
Relativity and Gravitation: Topical Collection".

\bibitem{bek} J. D. Bekenstein, Lett. Nuovo Cimento \textbf{11}, 467 (1974).

\bibitem{gib} G.~W.~Gibbons, Commun.\ Math.\ Phys.\ \textbf{44}, 245 (1975).

\bibitem{Bekenstein:1974ax} J.~D.~Bekenstein, 
Phys.\ Rev.\ D \textbf{9}, 3292 (1974).

\bibitem{Hawking:1974rv} S.~W.~Hawking, 
Nature \textbf{248}, 30 (1974).

\bibitem{Hawking} S.~W.~Hawking, 
Commun.\ Math.\ Phys.\ \textbf{43}, 199 (1975) Erratum: [Commun.\ Math.\
Phys.\ \textbf{46}, 206 (1976)].

\bibitem{Unruh:1976db} W.~G.~Unruh, 
Phys.\ Rev.\ D \textbf{14}, 870 (1976).

\bibitem{Parikh:1999mf} M.~K.~Parikh and F.~Wilczek, 
Phys.\ Rev.\ Lett.\ \textbf{85}, 5042 (2000).

\bibitem{Akhmedov:2006pg} E.~T.~Akhmedov, V.~Akhmedova and D.~Singleton, 
Phys.\ Lett.\ B \textbf{642}, 124 (2006).

\bibitem{Kerner:2006vu} R.~Kerner and R.~B.~Mann, 
Phys.\ Rev.\ D \textbf{73}, 104010 (2006).

\bibitem{Angheben:2005rm} M.~Angheben, M.~Nadalini, L.~Vanzo and S.~Zerbini, 
JHEP \textbf{0505}, 014 (2005).

\bibitem{Singleton:2011vh} D.~Singleton and S.~Wilburn, 
Phys.\ Rev.\ Lett.\ \textbf{107}, 081102 (2011).

\bibitem{Kubiznak:2012wp} D.~Kubiznak and R.~B.~Mann, 
JHEP \textbf{1207}, 033 (2012).

\bibitem{Kubiznak:2016qmn} D.~Kubiznak, R.~B.~Mann and M.~Teo, 
Class.\ Quant.\ Grav.\ \textbf{34}, no. 6, 063001 (2017).

\bibitem{Sakalli:2015taa} I.~Sakalli and A.~Ovgun, 
Eur.\ Phys.\ J.\ Plus \textbf{130}, no. 6, 110 (2015).

\bibitem{Sakalli:2014sea} I.~Sakalli and A.~Ovgun, 
EPL \textbf{110}, no. 1, 10008 (2015).

\bibitem{Sakalli:2010yy} I.~Sakalli, M.~Halilsoy and H.~Pasaoglu, 
Int.\ J.\ Theor.\ Phys.\ \textbf{50}, 3212 (2011)

\bibitem{Ovgun:2015jna} A.~\"{O}vg\"{u}n, 
Int.\ J.\ Theor.\ Phys.\ \textbf{55}, no. 6, 2919 (2016).

\bibitem{Sakalli:2016cbo} I.~Sakalli and A.~\"{O}vg\"{u}n, 
Eur.\ Phys.\ J.\ Plus \textbf{131}, no. 6, 184 (2016).

\bibitem{Sakalli:2017ewb} I.~Sakalli and A.~Ovgun, 
EPL \textbf{118}, no. 6, 60006 (2017).

\bibitem{Jusufi:2015zmv} K.~Jusufi, 
Gen.\ Rel.\ Grav.\ \textbf{48}, no. 8, 105 (2016).

\bibitem{Jusufi:2015eta} K.~Jusufi, 
Gen.\ Rel.\ Grav.\ \textbf{47}, no. 10, 124 (2015).

\bibitem{Kuang:2017sqa} X.~M.~Kuang, J.~Saavedra and A.~\"{O}vg\"{u}n, 
Eur.\ Phys.\ J.\ C \textbf{77}, no. 9, 613 (2017).

\bibitem{Born:1934ji} M.~Born, 
Proc.\ Roy.\ Soc.\ Lond.\ A \textbf{143}, no. 849, 410 (1934).

\bibitem{gross} D.~J.~Gross and J.~H.~Sloan, 
Nucl.\ Phys.\ B \textbf{291}, 41 (1987).

\bibitem{sezgin} E.~Bergshoeff, E.~Sezgin, C.~N.~Pope and P.~K.~Townsend, 
Phys.\ Lett.\ B \textbf{188}, 70 (1987).

\bibitem{dilatonBI} G. Panotopoulos and A. Rincon, Phys. Rev. D \textbf{96},
025009 (2017).

\bibitem{bibh2} R. Yamazaki and D. Ida, Phys. Rev. D \textbf{64}, 024009
(2001).

\bibitem{bibh3} A.~Sheykhi, N.~Riazi and M.~H.~Mahzoon, 
Phys.\ Rev.\ D \textbf{74}, 044025 (2006).

\bibitem{Vilenkin:1984ib} A.~Vilenkin, 
Phys.\ Rept.\ \textbf{121}, 263 (1985).

\bibitem{Vanchurin:2005pa} V.~Vanchurin, K.~D.~Olum and A.~Vilenkin, 
Phys.\ Rev.\ D \textbf{74}, 063527 (2006).

\bibitem{Polchinski:2007rg} J.~Polchinski and J.~V.~Rocha, 
Phys.\ Rev.\ D \textbf{75}, 123503 (2007).

\bibitem{Blanco-Pillado:2017oxo} J.~J.~Blanco-Pillado and K.~D.~Olum, 
Phys.\ Rev.\ D \textbf{96}, no. 10, 104046 (2017).

\bibitem{Blanco-Pillado:2017rnf} J.~J.~Blanco-Pillado, K.~D.~Olum and
X.~Siemens, 
Phys.\ Lett.\ B \textbf{778}, 392 (2018).

\bibitem{Jusufi:2017uhh} K.~Jusufi and A.~\"{O}vg\"{u}n, 
Phys. Rev. D \textbf{97}, 064030 (2018).

\bibitem{byform} J. D. Brown and J. W. York, Phys. Rev. D \textbf{47}, 1407
(1993).

\bibitem{PRLqlm} M-T. Wang, S-T. Yau, Phys. Rev. Lett. \textbf{102}, 021101
(2009).

\bibitem{fabris} G. Clement, J. C. Fabris, and G. T. Marques, Phys. Lett. B 
\textbf{65}, 54 (2007).

\bibitem{swald} R. M. Wald, \textit{General Relativity} (The University of
Chicago Press, Chicago and London, 1984).

\bibitem{bekent} J.D. Bekenstein, Lett. Nuovo Cimento \textbf{4}, 737 (1972).

\bibitem{bekent2} J.D. Bekenstein, Phys. Rev. D \textbf{7}, 2333\ (1973).

\bibitem{mkge} A. Ejaz, H. Gohar, H. Lin, K. Saifullah, and S. T. Yau, Phys.
Lett. B \textbf{726}, 827 (2013).

\bibitem{sf1} M. Abramowitz and I. A. Stegun, \textit{Handbook of
Mathematical Functions} (Dover, New York, 1965).

\bibitem{sf2} S. Y. Slavyanov and W. Lay, \textit{Special Functions: A
Unified Theory Based on Singularities}, (Oxford Mathematical Monographs, New
York, 2000).

\bibitem{heunc1} A. Ronveaux, \textit{Heun's differential equations},
(Oxford University Press, New York, 1995).

\bibitem{heunc2} R. S. Maier, \textit{The 192 Solutions of Heun Equation},
(Preprint math CA/0408317, 2004).

\bibitem{heunc3} P. P. Fiziev, J. Phys. A: Math. Theor. \textbf{43}, 035203
(2010).

\bibitem{heunc4} I. Sakalli and M. Halilsoy, Phys. Rev. D \textbf{69},
124012 (2004).

\bibitem{heunc5} A. Al-Badawi and I. Sakalli, Jour. Math. Phys. \textbf{49},
052501 (2008).

\bibitem{heunc6} I. Sakalli and A. Al-Badawi, Can. J. Phys. \textbf{87}, 349
(2009).

\bibitem{heunc7} M. Hortacsu, Adv. High Energy Phys. \textbf{2018}, 8621573
(2018).

\bibitem{heunc8} M. Hortacsu, \textquotedblleft \textit{Heun functions and
their uses in physics},\textquotedblright\ in \textit{Proceedings of the
13th Regional Conference on Mathematical Physics}, U. Camci and I. Semiz,
Eds., pp. 23--39, World Scientific, Antalya, Turkey, October, 2010.

\bibitem{heunc9} I. Sakalli, Phys. Rev. D \textbf{94}, 084040 (2016).

\bibitem{grg2018} K. Jusufi, I. Sakalli, and A. Ovgun, Gen Relativ Gravit 
\textbf{50}, 10 (2018).

\bibitem{waveid} A. L\'{o}pez-Ortega, Int. J. Mod. Phys. D \textbf{18}, 1441
(2009).

\bibitem{waveid2} A. L\'{o}pez-Ortega and I. Vega-Acevedo, Gen. Relativ.
Gravit. \textbf{43}, 2631 (2011).

\bibitem{uns1} M. Azreg-Ainou, Class. Quantum Gravity \textbf{16}, 245
(1999).

\bibitem{uns2} R. Becar, S. Lepe, and J. Saavedra, Phys. Rev. D \textbf{75},
084021 (2007).

\bibitem{epjc2018} K. Destounis, G. Panotopoulos, and A. Rincon, Eur. Phys.
J. C \textbf{78}, 139 (2018).

\bibitem{our1} E. Motl and A. Neitzke, Adv.Theor.Math.Phys. \textbf{7}, 307
(2003).

\bibitem{our2} M.~Casals and A.~C.~Ottewill, 
Phys.\ Rev.\ D \textbf{97}, no. 2, 024048 (2018).

\bibitem{our4} E.~Berti, V.~Cardoso, K.~D.~Kokkotas and H.~Onozawa, 
Phys.\ Rev.\ D \textbf{68}, 124018 (2003).

\bibitem{PRL2004} Ch. Buggle, J. Leonard, W. von Klitzing, and J.T.M.
Walraven, Phys. Rev. Lett. \textbf{93}, 173202 (2004).

\bibitem{dejan} R.~Dong, W.~H.~Kinney and D.~Stojkovic, 
JCAP \textbf{1610}, no. 10, 034 (2016).

\bibitem{HOD} S.~Hod, 
Phys.\ Rev.\ D \textbf{75}, 064013 (2007).

\bibitem{zhidenko} M.~A.~Cuyubamba, R.~A.~Konoplya and A.~Zhidenko, 
Phys.\ Rev.\ D \textbf{93}, no. 10, 104053 (2016).

\bibitem{vieira1} H.~S.~Vieira, V.~B.~Bezerra and A.~A.~Costa, 
EPL \textbf{109}, no. 6, 60006 (2015).

\bibitem{csp1} G. De A. Marques and V. B. Bezerra, \textit{
Some quantum effects in the spacetimes of topological defects},  The Tenth Marcel Grossmann Meeting. Proceedings of the MG10 Meeting held at Brazilian Center for Research in Physics (CBPF), Rio de Janeiro, Brazil, 20-26 July 2003, Eds.: Mário Novello; Santiago Perez Bergliaffa; Remo Ruffini. Singapore: World Scientific Publishing, in 3 volumes, ISBN 981-256-667-8 (set), ISBN 981-256-980-4 (Part A), ISBN 981-256-979-0 (Part B), ISBN 981-256-978-2 (Part C), Part C, p. 2199 - 2201 (2006).

\bibitem{csp2} E.~Hackmann, B.~Hartmann, C.~L\"{a}mmerzahl, and P.~Sirimachan, Phys. Rev. D \textbf{81}, 064016 (2010).

\bibitem{csp3} E.~Hackmann, B.~Hartmann, C.~L\"{a}mmerzahl, and P.~Sirimachan, Phys. Rev. D \textbf{82}, 044024 (2010). 


\bibitem{vieira2} H.~S.~Vieira, 
Chin.\ Phys.\ C \textbf{41}, no. 4, 043105 (2017).

\bibitem{corda1} C.~Corda, 
JHEP \textbf{1108}, 101 (2011).

\bibitem{corda2} C.~Corda, 
Int.\ J.\ Mod.\ Phys.\ D \textbf{21}, 1242023 (2012).

\bibitem{corda3} C.~Corda, 
Adv.\ High Energy Phys.\ \textbf{2015}, 867601 (2015).

\bibitem{alberto} C.~Chirenti, A.~Saa and J.~Skakala, 
Phys.\ Rev.\ D \textbf{87}, no. 4, 044034 (2013).

\bibitem{app1} A. Decarreau, M. C. Dumont-Lepage, P. Maroni, A. Robert, and
A. Ronveaux, \textit{Formes Canoniques de Equations confluentes de
l'equation de Heun}, Ann. Soc. Sci. Bruxelles. \textbf{T92(I-II)}, 53 (1978).

\bibitem{app2} See \textit{Maple\texttrademark\ 2018 (maplesoft.com)}.

\bibitem{app3} A. Decarreau, P. Maroni, and A. Robert, Ann. Soc. Sci.
Bruxelles. \textbf{T92(III)}, 151 (1978).

\bibitem{app4} P. P. Fiziev, \textit{Classes of Exact Solutions to
Regge-Wheeler and Teukolsky Equations}, arXiv:0902.1277 (2009).
\end{thebibliography}
\end{document}